\documentclass[sigconf]{aamas}

\usepackage{balance} 
\usepackage{algorithm}
\usepackage{algpseudocode}
\algrenewcommand{\algorithmiccomment}[1]{\hfill$\triangleright$ #1}

\setcopyright{ifaamas}
\acmConference[AAMAS '26]{Proc.\@ of the 25th International Conference
on Autonomous Agents and Multiagent Systems (AAMAS 2026)}{May 25 -- 29, 2026}
{Paphos, Cyprus}{C.~Amato, L.~Dennis, V.~Mascardi, J.~Thangarajah (eds.)}
\copyrightyear{2026}
\acmYear{2026}
\acmDOI{}
\acmPrice{}
\acmISBN{}

\acmSubmissionID{<<731>>}

\title[AAMAS-2026 Formatting Instructions]{GraphMASAL: A Graph-based Multi-Agent System for Adaptive Learning}

\author{Biqing Zeng}
\authornote{Corresponding author}
\affiliation{
  \institution{South China Normal University}
  \city{Foshan}
  \country{China}}
\email{zengbiqing137@163.com} 

\author{Mengquan Liu}
\affiliation{
  \institution{South China Normal University}
  \city{Foshan}
  \country{China}}
\email{2860251365@qq.com}

\author{Zongwei Zhen}
\affiliation{
  \institution{South China Normal University}
  \city{Foshan}
  \country{China}}
\email{1169619484@qq.com}

\begin{abstract}
The advent of Intelligent Tutoring Systems (ITSs) has marked a paradigm shift in education, enabling highly personalized learning pathways. However, true personalization requires adapting to learners’ complex knowledge states (multi-source) and diverse goals (multi-sink); existing ITSs often lack the necessary structural-reasoning capability and knowledge dynamism to generate genuinely effective learning paths, and they lack scientifically rigorous validation paradigms. In this paper we propose GraphMASAL (A Graph-based Multi-Agent System for Adaptive Learning), which integrates (i) a dynamic knowledge graph for persistent, stateful learner modeling; (ii) a LangGraph-orchestrated trio of agents (Diagnoser, Planner, Tutor); (iii) a knowledge-graph–grounded two-stage neural IR component (dual-encoder dense retrieval with cross-encoder listwise re-ranking and calibrated score fusion); and (iv) a multi-source multi-sink (MSMS) planning engine with a cognitively grounded cost and an approximation guarantee via greedy set cover.
Under blinded automated evaluations with matched inputs and inference settings across diverse student profiles, GraphMASAL consistently outperforms LLM prompting and structured ablations in planning—achieving stronger structural/sequence alignment of learning paths, higher coverage of weak concepts, and lower learning cost—while also surpassing prompt-based baselines in cognitive diagnosis. Agreement with expert/LLM-proxy ratings further supports the validity of our evaluation protocol.
These findings indicate that grounding LLM agents in a dynamic knowledge graph, coupled with optimization under educational constraints, yields reliable, interpretable, and pedagogically plausible learning plans, advancing personalized and goal-oriented education.

\end{abstract}


\newcommand{\BibTeX}{\rm B\kern-.05em{\sc i\kern-.025em b}\kern-.08em\TeX}

\begin{document}


\pagestyle{fancy}
\fancyhead{}


\maketitle 


\section{Introduction}

The advent of Intelligent Tutoring Systems (ITSs) has marked a paradigm shift in education, moving us closer to the long-held vision of a truly personalized learning experience for every individual~\cite{Alkhatlan2018, VanLehn2006}. The ultimate goal is an intelligent system that not only possesses a deep understanding of a dynamic knowledge domain but can also perform complex reasoning to execute multi-step, adaptive instructional tasks. Such a system would act as a personal mentor, guiding a learner from their current state of understanding towards their unique learning objectives.

However, the path to achieving this vision is fraught with systemic challenges that have prevented existing systems from reaching their full potential. We identify four interconnected bottlenecks that must be addressed in a holistic manner:

First is the Knowledge Layer challenge. In an era of rapid information growth, knowledge is not static. Most ITSs are built upon fixed curricula or static knowledge bases, which quickly become outdated and fail to capture the intricate, evolving relationships between concepts. A robust foundation for adaptive learning requires a dynamic knowledge graph that can continuously evolve, reflecting the current state of the domain.

Second, we face the Execution Layer challenge. The pedagogical process of "diagnose, plan, and tutor" is not a monolithic task but a complex workflow requiring distinct expertise. Entrusting this entire chain to a single, monolithic model or a simple series of large language model (LLM) calls leads to brittleness and a lack of specialized reasoning. True personalization requires a team of collaborative agents, each an expert in its sub-task, working in concert to create a seamless and effective learning experience.

Third, at the heart of this workflow lies the Decision Core challenge. Even with a perfect knowledge base and a team of expert agents, the fundamental task of charting an optimal learning path is a profound computational problem. Mapping a learner's initial knowledge state (a multi-source problem) to their diverse goals (a multi-sink problem) is an NP-hard optimization task. Existing agents lack the sophisticated "cognitive engine" needed to navigate this complexity and generate paths that are not just correct, but pedagogically efficient and sound.

Finally, towering over all these is the Scientific Validation challenge. If we build such a complex system, how do we scientifically measure the quality of its output? Evaluating a generated learning path is non-trivial. Simple accuracy metrics are insufficient; we need a new evaluation paradigm capable of assessing a path's logical coherence, pedagogical soundness, and structural integrity compared to an expert-defined ideal. Without this, the field cannot reliably measure progress.

To overcome these systemic barriers, we propose GraphMASAL, an integrated, graph-based multi-agent system for adaptive learning. Our framework is designed as a complete, end-to-end solution that systematically addresses each of the four challenges. The contributions of this paper are, therefore, four-fold:

We establish a dynamic knowledge graph as the system's foundation, which not only ensures the domain knowledge is current and comprehensive but also allows for real-time updates to reflect a student's evolving knowledge state.

We design a three-agent collaborative architecture (Diagnoser, Planner, Tutor) as the workflow engine, enabling modular, automated, and expert-driven execution of the tutoring process.

We introduce a novel multi-source, multi-sink path optimization algorithm as the core decision-making engine, empowering the Planner agent with the intelligence to generate highly effective learning routes.

We propose a new scientific evaluation paradigm, centered on our PathSim metric, to provide a robust and reliable methodology for measuring the quality of the system's output.
Beyond these, we introduce a KG-enhanced semantic retrieval pipeline with principled ranking fusion, and provide a formal treatment of the multi-source multi-sink (MSMS) planning with approximation guarantees under a cognitively grounded cost model.

This paper details the architecture of GraphMASAL, elaborates on its core algorithmic and methodological innovations, and presents extensive experimental validation demonstrating its superiority over existing approaches.


\section{RELATED WORKS}

\subsection{\textbf{The Evolution of Personalized Learning and Its Systemic Challenges}}
Personalized learning in ITS seeks to scale one-to-one tutoring by constructing precise learner and content models~\cite{VanLehn2006,Abyaa2019,Sottilare2013}, evolving from early rule-based systems to modern data-driven platforms that leverage machine learning~\cite{Alkhatlan2018}. Despite progress, deep personalization remains constrained by coupled challenges in knowledge dynamism, execution complexity, algorithmic optimization, and scientific validation. GraphMASAL is designed to address these challenges holistically.

\subsection{\textbf{Knowledge Representation for Adaptive Learning}}
A system's ability to reason and personalize is fundamentally constrained by its method of knowledge representation. The field has evolved from static, schema-oriented ontologies—which, despite their rich semantics, are ill-suited for rapidly changing domains—to more flexible, instance-oriented Knowledge Graphs (KGs)~\cite{Wang2021KG}. Unlike concept or mind maps designed for human cognition, KGs are machine-readable structures that provide a robust foundation for advanced AI tasks.

\subsection{\textbf{Modern Agentic Workflows with LLMs and LangGraph}}

The emergence of Large Language Models (LLMs) has revolutionized the creation of intelligent agents~\cite{Wang2024LLM}. However, orchestrating multiple LLM-based agents to perform complex tasks introduces significant challenges in state management, reliable communication, and workflow control~\cite{Nye2023}. To address this, a new generation of agentic frameworks, such as LangGraph, has emerged. LangGraph enables the modeling of complex, stateful, multi-agent applications as graphs, where nodes represent agents or tools and edges define the control flow~\cite{Guo2024}. This provides a modular, transparent, and persistent structure that resolves the rigidity and state-management issues of earlier multi-agent systems.

GraphMASAL's three-agent architecture (Diagnoser, Planner, Tutor) is an advanced application of this modern paradigm. By implementing this classic ITS role division within LangGraph's robust orchestration framework, our system ensures a seamless and logically sound flow of information. This approach mirrors the evolution in software engineering from monolithic architectures to orchestrated microservices~\cite{Qian2024, Park2023}, bringing a new level of scalability, maintainability, and extensibility to the design of Intelligent Tutoring Systems. 

Beyond orchestration, our retrieval stack follows the dense-retrieval + cross-encoder reranking paradigm popular in neural IR and learning-to-rank~\cite{Lewis2020}. Different from generic RAG, we ground dense retrieval on KG nodes and prerequisite subgraphs, enabling structure-aware reranking and personalization.

\begin{figure*}[t]
  \centering
  \includegraphics[width=0.9\textwidth]{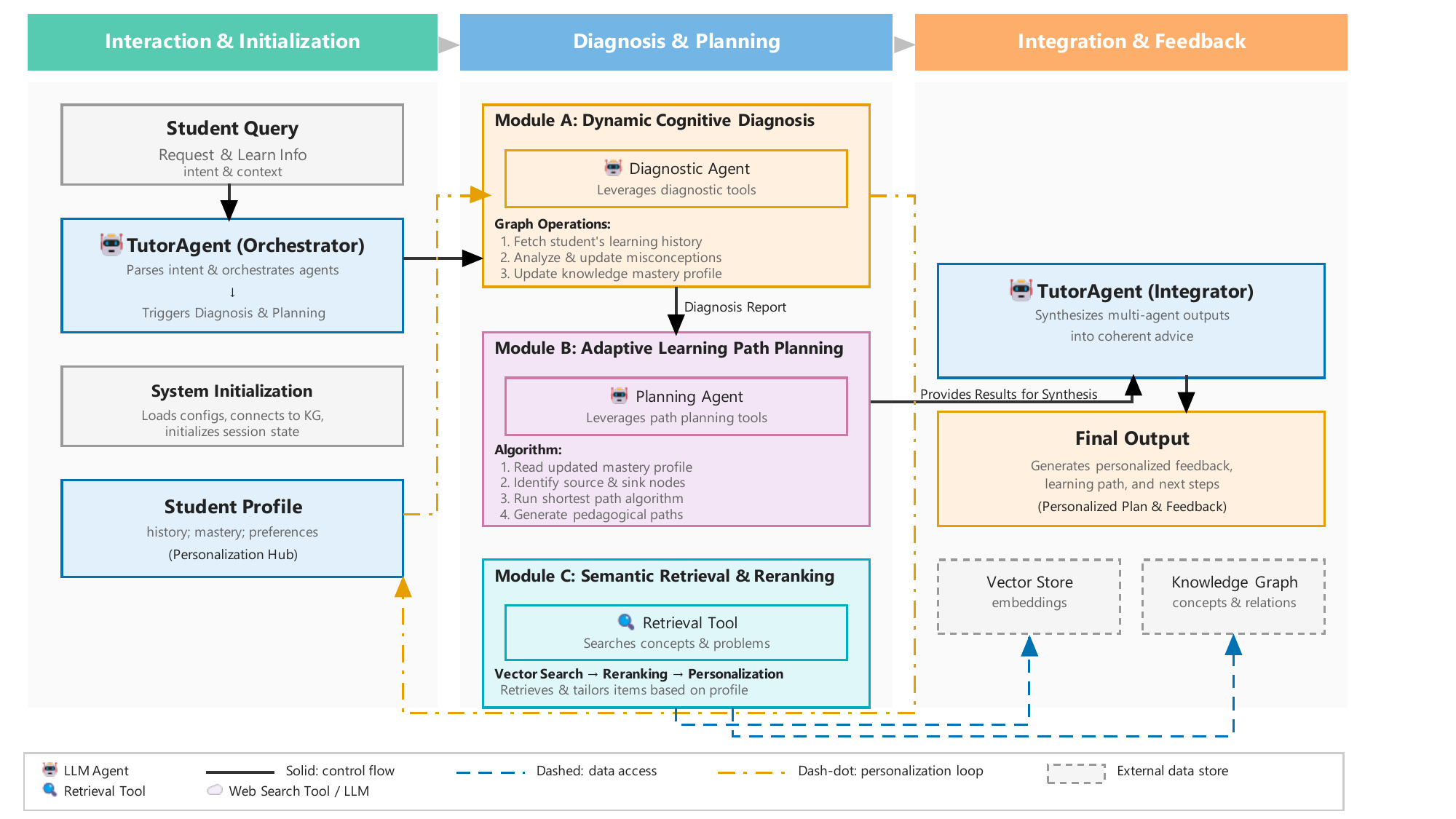}
  \caption{System architecture of GraphMASAL showing three phases with modular agent components and data flows.}
  \label{fig:system_architecture}
  \Description{A flowchart illustrating the system architecture of GraphMASAL. It begins with User Input in Phase 1, proceeds to Phase 2 with three modules (Dynamic Cognitive Diagnosis, Adaptive Learning Path Planning, and Semantic Retrieval), and concludes with Phase 3 for Integration & Feedback. A foundation layer with a Knowledge Graph and other services supports the entire process.}
\end{figure*}


\section{The Proposed Framework}

To achieve truly personalized, dynamic, and effective adaptive learning, we designed and implemented GraphMASAL, a graph-based multi-agent system. The core principle of this framework is to unify the modeling of a student's cognitive state, the domain knowledge structure, and pedagogical strategies within a dynamically evolving environment, driven by the collaborative reasoning of intelligent agents. his section elaborates on the framework's four core system components—the semantic retrieval engine for knowledge grounding, the dynamic knowledge graph as the knowledge foundation, the multi-agent collaborative workflow, and the path optimization algorithm as the decision engine—as well as the PathSim methodology for scientific evaluation.

\subsection{Semantic Retrieval Engine and Personalization}
We operationalize a KG-enhanced retrieval pipeline with four stages: (1) \textbf{Embedding} texts (queries, concept names/descriptions, item stems) using \texttt{doubao-embedding-text-240715}; (2) \textbf{Vector Similarity} search over concept/item vectors (cosine TopK) backed by Neo4j indices; (3) \textbf{Cross-Encoder Reranking} to re-score candidates with context-aware matching; and (4) \textbf{Personalization} that annotates candidates with student-specific mastery and misconceptions. Given a query $q$ and candidate $d$, we fuse bi-encoder similarity and cross-encoder scores:

\begin{equation}
\text{Score}(d \mid q) 
= \alpha\, \cos\!\big(\mathbf{e}(q), \mathbf{e}(d)\big) 
+ (1-\alpha)\, \sigma\!\big(f_{\text{ce}}(q,d)\big),
\end{equation}
where $\mathbf{e}(\cdot)$ is the embedding, $f_{\text{ce}}$ is the cross-encoder, $\sigma$ is a min-max normalization over the candidate set, and $\alpha \in [0,1]$ (default $\alpha{=}0.5$). We return Top-$K$ (default $K{=}6$) to the Tutor agent or tools (e.g., \texttt{QueryTool}).

\begin{algorithm}[b]
\caption{KG-Enhanced Semantic Retrieval with Reranking}
\label{alg:retrieval}
\begin{algorithmic}[1]
\State Input: query $q$, candidate set $\mathcal{D}$, weights $\alpha$, cutoffs $K$
\State Embed $q \!\to\! \mathbf{e}(q)$; for each $d\!\in\!\mathcal{D}$, get $\mathbf{e}(d)$
\State Compute $\text{sim}(q,d) = \cos(\mathbf{e}(q), \mathbf{e}(d))$; keep Top-$K_v$
\State Re-rank $\mathcal{D}_{K_v}$ using cross-encoder $f_{\text{ce}}(q,d)$; min-max normalize
\State Fuse scores by Eq.(1); return Top-$K$ with personalization annotations
\end{algorithmic}
\end{algorithm}

The system architecture of GraphMASAL, depicted in Figure 1, is a hierarchical and collaborative structure built upon a dynamic knowledge graph foundation that supports three specialized agents (Diagnoser, Planner, Tutor) orchestrated via LangGraph, with planning driven by the MSMS algorithm and validated through the PathSim evaluation paradigm.

\subsection{The Knowledge Foundation: Dynamic Knowledge Graph}
\begin{figure*}[thb]
  \centering
  \includegraphics[width=\linewidth]{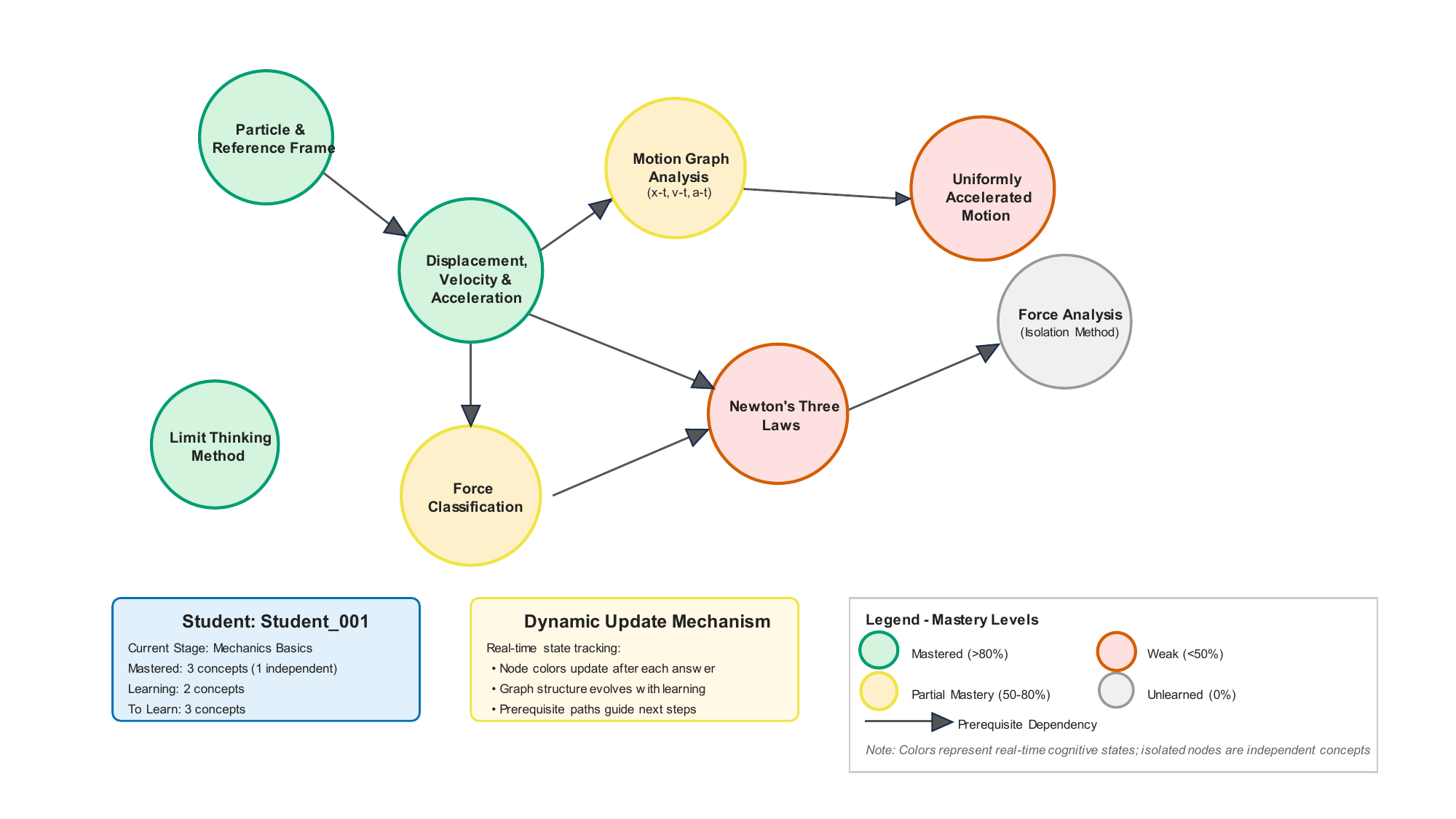}
  \caption{Dynamic Knowledge Graph Structure Physics Concept Network with Prerequisite Dependencies}
  \label{fig:system_architecture}
  \Description{A flowchart illustrating the system architecture of GraphMASAL. It begins with User Input in Phase 1, proceeds to Phase 2 with three modules (Dynamic Cognitive Diagnosis, Adaptive Learning Path Planning, and Semantic Retrieval), and concludes with Phase 3 for Integration & Feedback. A foundation layer with a Knowledge Graph and other services supports the entire process.}
\end{figure*}
The fundamental prerequisite for adaptive learning is the precise and real-time capture of a learner's dynamic cognitive state. Static knowledge bases cannot meet this requirement. Therefore, we implement a dynamic knowledge graph mechanism that supports real-time updates and structural evolution. Its "dynamic" nature is manifested on two levels:

\textbf{Structural Dynamics:} The knowledge graph's schema can be flexibly extended with the introduction of new courses and concepts.

\textbf{State Dynamics:} More critically, the attributes and relationships representing student states are updated in real-time with every student interaction (e.g., correctly or incorrectly answering a question).

This design transforms the knowledge graph from a static "knowledge map" into a dynamic "cognitive model," providing a solid and instantaneous data foundation for the decision-making of the higher-level agents, which is key to achieving truly personalized education.

\subsection{The Agentic Workflow: Multi-Agent Collaboration}
We decompose the complex pedagogical task into three collaborative agents and orchestrate their interactions using the LangGraph framework to ensure the stability, orderliness, and state consistency of the information flow.

Diagnostic Agent: Its core responsibility is "deep attribution"~\cite{Piech2015}. Upon receiving a student's attempt record, it not only judges correctness but, more importantly, queries the knowledge graph to associate the student's incorrect options with specific Misconception nodes and updates the mastery level of related Concept nodes accordingly. It answers the question, "Why did the student make this mistake?"

Planning Agent: Its core responsibility is "path optimization"~\cite{Li2021, Yun2023}. It receives the student's current cognitive state (i.e., a set of weak concepts) from the Diagnostic Agent and uses this as input to invoke its core planning engine (see Section 2.4) to generate one or more optimal learning paths. It answers the question, "What should the student learn next, and how?"

Tutor Agent: Its core responsibility is "interaction and coordination." It acts as the master controller, managing multi-turn dialogues with the user, dispatching requests to the Diagnostic or Planning Agents, and integrating their structured outputs to generate insightful and easy-to-understand natural language feedback.

By constructing a StateGraph with LangGraph, we can precisely define the trigger conditions and execution logic for each agent, making the entire collaborative process resemble a sophisticated "cognitive processing pipeline" that guarantees the coherence and logicality of the tutoring service.

\subsection{The Core Planning Engine: Multi-Source Multi-Sink Path Optimization Algorithm}
The technical core of our system is the Multi-Source Multi-Sink (MSMS) Path Optimization Algorithm embedded within the Planning Agent. Traditional pathfinding algorithms (single-source, single-sink) are ill-equipped to handle the complex, realistic scenario where a student has multiple knowledge gaps (multi-sink) and a heterogeneous knowledge base (multi-source). The MSMS algorithm is specifically designed to address this challenge.

The algorithm models the learning path generation task as an optimization problem on a graph. The set of "sources" comprises the concepts the student has already mastered, while the set of "sinks" represents the weak concepts to be learned. The objective is to find a set of paths originating from the sources that covers all sinks while minimizing the total learning cost. 

Crucially, the design preference of this algorithm is deeply inspired by theories from educational psychology and cognitive science. Specifically, its optimization objective—minimizing the total number of new concepts in the resulting paths—is intended to proactively manage the learner's Cognitive Load~\cite{Sweller2011}. By avoiding the introduction of excessive irrelevant or redundant concepts at once, we can reduce extraneous cognitive load, allowing the learner to allocate more cognitive resources to the internalization and construction of knowledge (germane load). Furthermore, the algorithm's "multi-source" nature, starting from what the student already knows, aligns with Constructivist Learning Theory~\cite{VanMerrienboer2017}. It scaffolds new learning by activating and connecting to the student's existing cognitive schemata, which is posited to enhance learning motivation and self-efficacy.

The detailed procedure is formalized in Algorithm 2.

\begin{figure*}[tbp]
  \centering
  \includegraphics[width=\linewidth]{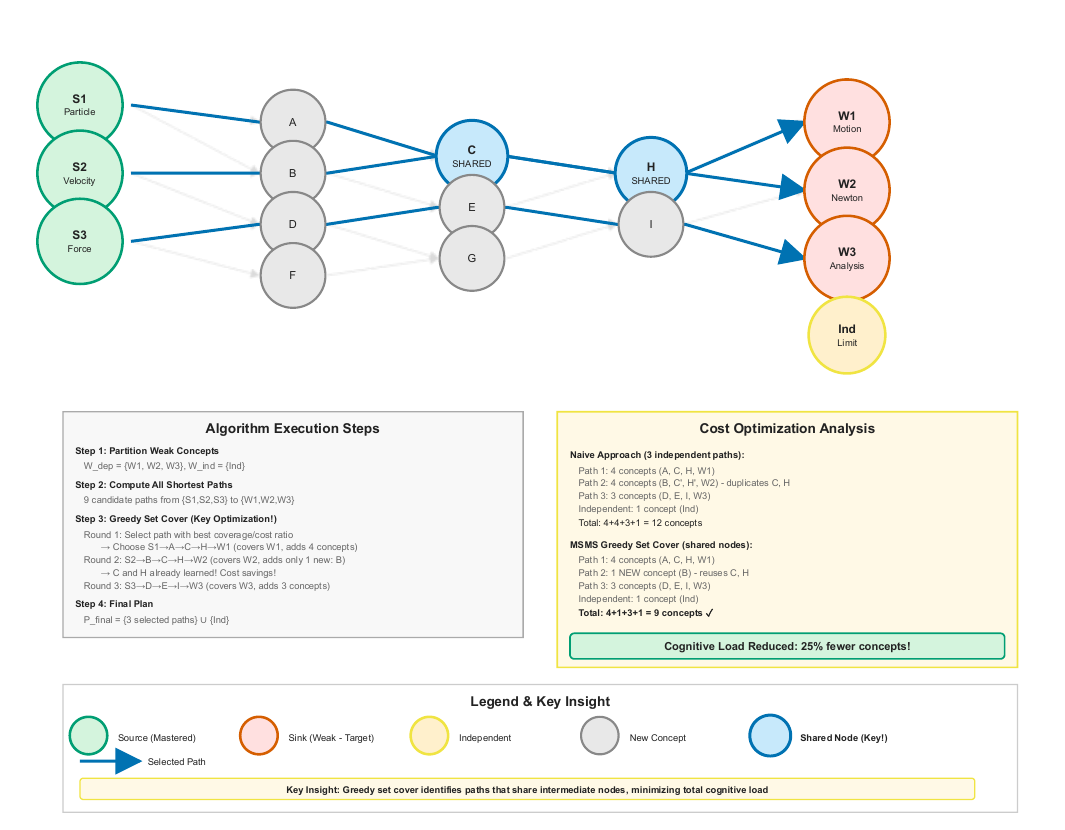}
  \caption{MSMS Algorithm for Multi-Source Multi-Sink Path Optimization via Greedy Set-Cover: Minimizing the Total Number of Emerging Concepts}
  \label{fig:system_architecture}
  \Description{Multi-Source Multi-Sink Path Optimization with Greedy Set Cover}
\end{figure*}

\begin{algorithm}[tbp]
\caption{MSMS Path Optimization}
\label{alg:msms}
\begin{algorithmic}[1]
\State \textbf{Input:}
\State $G = (V, E)$ \Comment{Knowledge graph where V are concepts and E are prerequisite edges}
\State $S \subset V$ \Comment{Set of mastered concepts (sources)}
\State $W \subset V$ \Comment{Set of weak concepts (sinks)}
\State
\State \textbf{Output:}
\State $P_{\text{final}}$ \Comment{A final learning plan, composed of paths and independent concepts}
\State
\Function{MsmsPlanner}{$G, S, W$}
    \State \textit{$\triangleright$ Step 1: Partition weak concepts}
    \State $W_{\text{ind}} \gets \{w \in W \mid \neg\exists w' \in W, (w', w) \in E\}$ \Comment{Independent concepts}
    \State $W_{\text{dep}} \gets W \setminus W_{\text{ind}}$ \Comment{Dependent concepts}
    \State
    \State \textit{$\triangleright$ Step 2: Compute all-pairs shortest paths from sources to dependent sinks}
    \State $\textit{AllPaths} \gets \emptyset$
    \ForAll{$s \in S$}
        \ForAll{$w \in W_{\text{dep}}$}
            \State $\textit{path} \gets \text{DIJKSTRA}(G, s, w)$
            \If{\textit{path} exists}
                \State Add \textit{path} to \textit{AllPaths}
            \EndIf
        \EndFor
    \EndFor
    \State
    \State \textit{$\triangleright$ Step 3: Greedily select paths to cover all dependent sinks}
    \State $P_{\text{optimal}} \gets \text{GREEDY\_SET\_COVER}(\textit{AllPaths}, W_{\text{dep}})$
    \State
    \State \textit{$\triangleright$ Step 4: Combine results into the final plan}
    \State $P_{\text{final}} \gets (P_{\text{optimal}}, W_{\text{ind}})$
    \State \Return $P_{\text{final}}$
\EndFunction
\end{algorithmic}
\end{algorithm}

\subsubsection{Problem Formulation and Cognitive Cost Model}
Let $G=(V,E)$ be the prerequisite graph over concepts. For a given student, $S\subset V$ denotes mastered sources and $W\subset V$ weak sinks. A feasible plan selects paths $\mathcal{P}$ such that each $w\in W$ is covered by at least one path $P: s\to w$ with $s\in S$. We minimize the total \emph{learning cost}
\begin{equation}
\min_{\mathcal{P}} \; \sum_{v \in \cup_{P\in\mathcal{P}} P} c(v)
\quad \text{s.t. } \forall w\in W,\; \exists P\in \mathcal{P}: P: s\to w,
\end{equation}
where $c(v)=\lambda_1\big(1-\text{mastery}(v)\big)+\lambda_2\,\text{difficulty}(v)+\lambda_3\,\text{fanout}(v)$. The cost penalizes novelty and split-attention while favoring scaffolding from mastered concepts.

\paragraph{Operational definitions and normalization.}

We instantiate the cost terms on the prerequisite graph as follows, with all components min–max normalized to $[0,1]$ for scale comparability.

\emph{mastery}$(v)$ is the student-specific proficiency stored in edge attribute \texttt{proficiency}$\in[0,1]$ of relation \texttt{HAS\_MASTERY\_OF}. 
\emph{difficulty}$(v)$ is a concept-level proxy derived as the normalized mean difficulty of problems linked to $v$ via \texttt{TESTS\_CONCEPT}; when unavailable, we use normalized \texttt{Concept.level}. 
\emph{fanout}$(v)$ quantifies local branching computed as normalized out-degree $\deg^{\text{out}}(v)$ under \texttt{IS\_PREREQUISITE\_FOR}. 
This design penalizes detours through highly branching regions while encouraging coherent, scaffolded progress.

\subsubsection{Complexity and Approximation Guarantee}
MSMS planning generalizes Set Cover: each candidate path covers a subset of sinks. Minimizing total node cost to cover all sinks is NP-hard by reduction. Our planner enumerates shortest source$\to$sink paths (depth-limited) and applies greedy set-cover selection over path-sets. Let $|W|$ be the number of sinks; the greedy procedure achieves a $(1+\ln |W|)$ approximation to the optimal cover cost~\cite{Chvatal1979SetCover,Slavik1997Greedy}. Bounding path length and pruning by prerequisite depth control enumeration cost while preserving high-quality candidates.

\subsubsection{Pedagogical Validity}
The cost $c(\cdot)$ operationalizes cognitive load theory: novelty and difficulty encode intrinsic/extraneous load; fan-out mitigates split attention. Multi-source anchoring aligns with constructivism by tying new learning to mastered schemata, yielding plans that are both computationally principled and pedagogically meaningful.

\subsection{The Scientific Evaluation Paradigm: PathSim Methodology}
To scientifically and objectively evaluate the performance of our planning algorithm, we propose a novel path similarity evaluation methodology, PathSim. Traditional graph or set similarity metrics (e.g., Jaccard similarity) are severely limited when assessing learning paths because they ignore two crucial pieces of information:

The Topological Structure of a Path: A learning path is more than a set of concepts; its internal node sequence and connectivity (i.e., dependency relations) are paramount.

The Composite Structure of a Plan: A complete learning plan is often a hybrid of multiple paths and several standalone concepts.

The PathSim method, a Topology-Aware Hybrid Path Matching Algorithm, is designed to address this gap. It employs a multi-level comparison framework that separately computes the similarity of the independent concept sets (using Jaccard similarity) and the path sets. When calculating path similarity, it holistically considers the overlap of nodes and edges, as well as the sequence similarity (calculated via normalized Levenshtein distance), thus enabling a more accurate measurement of the structural and sequential alignment between two paths.
The PathSim formalized in Algorithm 3.
\begin{algorithm}[tbp]
\caption{PathSim Similarity Calculation}
\label{alg:pathsim}
\begin{algorithmic}[1]
\State \textbf{Input:}
\State $Plan_A = (P_A, I_A)$ \Comment{Learning plan A}
\State $Plan_B = (P_B, I_B)$ \Comment{Learning plan B}
\State $w_p, w_i$ \Comment{Weights for path and independent concept similarity}
\State
\State \textbf{Output:}
\State $\textit{TotalSim}$ \Comment{The total similarity score between Plan\_A and Plan\_B}
\State
\Function{PathSim}{$Plan_A, Plan_B$}
    \State \textit{$\triangleright$ Step 1: Calculate similarity for independent concepts}
    \State $Sim_i \gets \text{JACCARD}(I_A, I_B)$
    \State
    \State \textit{$\triangleright$ Step 2: Calculate similarity for path sets}
    \If{$|P_A| = 0$ \textbf{and} $|P_B| = 0$}
        \State $Sim_p \gets 1$
    \ElsIf{$|P_A| = 0$ \textbf{or} $|P_B| = 0$}
        \State $Sim_p \gets 0$
    \Else
        \State \textit{$\triangleright$ 2a: Compute cross-similarity matrix M}
        \For{$i \gets 1 \text{ to } |P_A|$}
            \For{$j \gets 1 \text{ to } |P_B|$}
                \State $M[i, j] \gets \text{COMPUTE\_PATH\_SIM}(P_A[i], P_B[j])$ \Comment{Considers nodes, edges, sequence}
            \EndFor
        \EndFor
        \State
        \State \textit{$\triangleright$ 2b: Calculate symmetric average best match}
        \State $Sim_{A \to B} \gets \frac{1}{|P_A|} \sum_{i=1}^{|P_A|} \max_{j}(M[i, j])$
        \State $Sim_{B \to A} \gets \frac{1}{|P_B|} \sum_{j=1}^{|P_B|} \max_{i}(M[i, j])$
        \State $Sim_p \gets (Sim_{A \to B} + Sim_{B \to A}) / 2$
    \EndIf
    \State
    \State \textit{$\triangleright$ Step 3: Combine scores for the final similarity}
    \State $\textit{TotalSim} \gets w_p \cdot Sim_p + w_i \cdot Sim_i$
    \State \Return $\textit{TotalSim}$
\EndFunction
\end{algorithmic}
\end{algorithm}

The introduction of PathSim provides an effective and interpretable scientific paradigm for the evaluation and comparison of complex, structured learning plans, filling a void in existing evaluation methodologies.

\section{EXPERIMENTS}
To evaluate the effectiveness of GraphMASAL's output items, we conduct both LLM-based automated and human evaluations~\cite{Zheng2023}.
\subsection{Implementation and Experimental Setup}

\textbf{System}. We implement GraphMASAL in Python. Neo4j serves as the persistent graph store, supporting dynamic graph operations including node insertion, edge pruning, and relation rewiring under agent control. Workflow orchestration is realized with LangGraph~\cite{Guo2024}. All inference is powered by a commercial large language model, with doubao-seed-1.6-250615 as the reasoning backbone. For retrieval and entity grounding over the knowledge graph~\cite{Lewis2020}, we use doubao-embedding-text-240715 to index concept nodes, prerequisite relations, and item texts, enabling vector similarity search and semantic-level knowledge retrieval. We parameterize $\alpha{=}0.5$, $K{=}6$, and enumerate shortest paths up to length 10 before greedy selection; all hyper-parameters are released with our code.

\textbf{Data}. Our primary experimental assets consist of a curated high-school physics knowledge graph derived from publicly available textbooks: 101 concept nodes, 168 prerequisite edges, and 200 exercises drawn from the open-source multi-modal, multi-subject problem set \verb|mllm_multi_subject_data|. Each exercise is linked to one or more concept nodes and is accompanied by a misconception map enumerating common distractors and their underlying misunderstandings.

\textbf{Knowledge graph construction}. The knowledge graph is constructed from publicly available textbooks through an automated extraction pipeline. Concept nodes and prerequisite dependencies are extracted using doubao-seed-1.6-250615 with Chain-of-Thought prompting (CoTPrompt)~\cite{Wei2022}, which guides the model to identify domain concepts and their logical prerequisites through structured reasoning. To ensure graph quality and efficiency, we apply deduplication over semantically similar concepts based on embedding cosine similarity (threshold $>0.9$), yielding a concise, non-redundant knowledge structure that supports efficient path planning and retrieval.

\textbf{Simulated students}. To obtain controlled, repeatable evaluations, we synthesize 15 distinct student profiles. Each profile specifies an initial mastery vector over the 101 concepts. For every profile, we simulate a 15-step answering sequence. At each step, the correctness probability is sampled as a monotone function of the student’s current mastery over the exercise-linked concepts, thereby inducing individualized learning trajectories with realistic variance.

\textbf{Hyper-parameters and infrastructure}. temperature is set to 0.2 and \verb|top_p| to 0.9 for deterministic reasoning. The context window accommodates the retrieved top-k knowledge snippets (k = 6), the agent’s hidden chain-of-thought summary, and the tool outputs. All experiments are executed on a single workstation with standard CPU and GPU resources; retrieval and graph operations are I/O-bound and do not require specialized accelerators.
\subsection{Automated Evaluation}
Evaluating tutoring systems is challenging due to the open-ended nature of the tasks. To address this, we designed a scalable, automated evaluation pipeline based on quantitative metrics, which is then validated against an expert proxy.

\subsubsection{Overall Experimental Setup}
Our core evaluation methodology is a blind reasoning test. Crucially, we define the ground truth as an \emph{oracle baseline}—not an external gold standard from human experts, but rather the output of our own MSMS planner running with complete metadata access. Specifically, for each simulated student attempt, we execute the "teacher" version of GraphMASAL that has access to all problem annotations (\verb|linked_kp_ids| and \verb|misconception_map|), thereby producing the optimal diagnosis and learning path under perfect information. We then create a "blinded" version by removing these critical fields, forcing the deployed GraphMASAL to perform genuine reasoning and retrieval based solely on problem content. This setup measures how well our system reconstructs the oracle decision when key information must be inferred rather than directly retrieved.

\subsubsection{Evaluating Cognitive Diagnosis}
To rigorously evaluate the cognitive diagnosis capability of GraphMASAL, we conducted a systematic comparison against two representative baseline approaches: (1) Direct Prompting (DirPrompt), wherein the language model is tasked with diagnosing student misconceptions through straightforward task instructions, and (2) Chain-of-Thought Prompting (CoTPrompt)~\cite{Wei2022}, which incorporates explicit reasoning chains into the prompt to guide the model's diagnostic process. Both baselines leverage the same underlying large language model but differ in their prompting strategies. The diagnostic task requires the system to identify which concepts a student has mastered based on their problem-solving behaviors, a fundamentally challenging classification problem due to the fine-grained nature of concept-level attribution and the limited observable evidence from each student's answer history.
Our experimental results, as shown in Table~\ref{tab:final_results}, reveal a substantial performance gap between GraphMASAL and the prompt-based baselines. GraphMASAL significantly outperforms both DirPrompt and CoTPrompt across precision, recall, and F1-score metrics. This marked improvement demonstrates that GraphMASAL's integration of a dynamic knowledge graph, specialized diagnostic agent, and structured graph operations provides a more robust and reliable foundation for cognitive diagnosis than purely prompt-engineering approaches, which lack persistent memory of student states and struggle to reason over complex prerequisite dependencies.

\begin{table}[t]
\centering
 \caption{Comparison of evaluation results for Cognitive Diagnosis.}
 \label{tab:final_results}
\begin{tabular}{ll|ccc}
\toprule
\multicolumn{2}{c|}{\textbf{System}} & \textbf{Precision} & \textbf{Recall} & \textbf{F1 Score} \\
\midrule
& DirPrompt & 0.46 & 0.57 & 0.51 \\
& CoTPrompt & 0.54 & 0.60 & 0.57 \\
\midrule
& \textbf{GraphMASAL} & \textbf{0.80} & \textbf{0.69} & \textbf{0.74} \\
& \textit{w/o KG} & 0.62 & 0.64 & 0.63 \\

\bottomrule
 \end{tabular}
\end{table}

\subsubsection{Evaluating Learning Paths}
We benchmark path planning against LLM prompting and structured ablations using three metrics: PathSim (\textuparrow; structural/sequence alignment), Coverage (\textuparrow; proportion of weak concepts covered), and Total Cost (\textdownarrow; unique concepts to learn). For each of 15 simulated student profiles, all methods are evaluated under matched inputs (identical weak-concept set/targets, retrieval Top-$K$, and context window) and matched inference settings (same LLM backbone, fixed \texttt{temperature} and \texttt{top\_p}). To mitigate stochasticity, every configuration is repeated with 5 random seeds; we report mean $\pm$ 95\% CI over seeds\,$\times$\,profiles. Statistical significance is assessed via paired two-sided tests on per-profile scores.
MSMS (ours) attains the highest PathSim and Coverage while minimizing Total Cost, with improvements over all baselines that are statistically significant under paired tests in most profiles. \emph{Shortest-per-sink} achieves near-perfect Coverage but incurs higher cost and lower PathSim due to lack of global de-duplication. Removing the branching penalty (\emph{No-fanout}) increases redundancy, raising cost and slightly lowering PathSim. \emph{Bi-encoder only} outperforms \emph{w/o KG} yet lags MSMS without reranking/global optimization. \emph{w/o KG} and \emph{Random Path} yield the lowest alignment and coverage, with the latter exhibiting the highest cost. \emph{CoT Prompt} forms reasonable paths but underperforms MSMS on structure alignment and cost control.

\begin{table}[tb]
\centering
\caption{Path planning baselines: PathSim (\textuparrow), Coverage (\textuparrow), Total Cost (\textdownarrow).}
\label{tab:pathsim_baselines}
\begin{tabular}{lccc}
\toprule
\textbf{Method} & \textbf{PathSim} & \textbf{Coverage} & \textbf{Total Cost} \\
\midrule
\multicolumn{4}{l}{\textit{LLM-based}} \\
CoT Prompt & $0.71 \,\pm\, 0.12$ & $0.82$ & $16.6 \,\pm\, 4.2$ \\
\midrule
\multicolumn{4}{l}{\textit{Structured / Ablations }} \\
w/o KG & $0.63 \,\pm\, 0.15$ & $0.78$ & $18.3 \,\pm\, 4.7$ \\
Bi-encoder only & $0.74 \,\pm\, 0.10$ & $0.88$ & $15.5 \,\pm\, 3.9$ \\
Random Path & $0.52 \,\pm\, 0.18$ & $0.76$ & $21.9 \,\pm\, 5.6$ \\
Shortest-per-sink & $0.79 \,\pm\, 0.08$ & $0.96$ & $17.8 \,\pm\, 3.1$ \\
No-fanout penalty & $0.81 \,\pm\, 0.09$ & $0.97$ & $18.6 \,\pm\, 3.3$ \\
\midrule
\multicolumn{4}{l}{\textit{Proposed}} \\
\textbf{MSMS (ours)} & $\mathbf{0.857} \,\pm\, 0.07$ & $\mathbf{0.98}$ & $\mathbf{14.2} \,\pm\, 2.8$ \\
\bottomrule
\end{tabular}
\end{table}

\subsection{Human Validation of the Automated Evaluation}
To validate the reliability of our automated evaluation framework, we conducted a correlation analysis between automated metrics and human expert judgments. We randomly sampled 50 diagnosis-planning pairs from the experimental dataset. For each pair, three experienced physics educators independently rated the quality on a 5-point Likert scale, with ratings averaged as the human judgment baseline.

 Table~\ref{tab:human_validation} presents the Pearson correlation coefficients. Both diagnostic and planning quality show moderate-to-strong positive correlations with human ratings, with statistical significance at $p < 0.01$ and $p < 0.001$ respectively. Critically, human validation confirms that the oracle baseline itself—the idealized plan under perfect information—possesses pedagogical validity. This forms a \emph{logical closure}: our system reliably generates paths that closely match the oracle baseline, and human experts independently judge these oracle-aligned paths to be of high quality. Together, these findings validate not only the soundness of our automated metrics but also the pedagogical merit of the decision strategy our system consistently recovers.

\begin{table}[h]
\centering
\caption{Correlation between human expert ratings and automated metrics (N=50).}
\label{tab:human_validation}
\begin{tabular}{l|cc}
\toprule
\textbf{Evaluation Task} & \textbf{Correlation (r)} & \textbf{p-value} \\
\midrule
Diagnostic Quality & 0.65 & $< 0.01$ \\
Planning Quality & \textbf{0.68} & $< 0.001$ \\
\bottomrule
\end{tabular}
\end{table}

\section{Conclusion}

This work introduced GraphMASAL, a graph-centric multi-agent ITS that integrates: (i) a dynamic knowledge graph for persistent, stateful learner modeling; (ii) a LangGraph-based orchestration layer enabling interpretable, tool-augmented collaboration among Diagnoser, Planner, and Tutor agents; (iii) a KG-grounded two-stage neural IR component (dual-encoder dense retrieval with cross-encoder listwise re-ranking and calibrated score fusion) for concept and item grounding; and (iv) a multi-source multi-sink (MSMS) planning engine with a cognitively grounded cost and a $(1+\ln|W|)$ approximation via greedy set cover over source$\to$sink paths. Across blinded evaluations and expert validations, GraphMASAL yielded substantially higher diagnostic fidelity (F1 $=0.74$) and stronger structural alignment of learning plans (mean PathSim $=0.857$) with positive correlations to human judgments (diagnosis $r{=}0.65$, planning $r{=}0.68$), indicating both computational effectiveness and pedagogical plausibility. Beyond empirical gains, our formulation clarifies how explicit graph structure, optimization under educational constraints, and modular agent tooling jointly contribute to reliability and interpretability in ITS. Future work includes principled incorporation of multimodal resources into the KG, tighter coupling of affect/motivation signals with planning objectives, and longitudinal field deployments to quantify learning gains and equity across populations.




\newpage
\bibliographystyle{ACM-Reference-Format} 
\bibliography{sample}


\end{document}